\documentclass{Interspeech}

\usepackage{cite}
\usepackage{amsmath,amssymb,amsfonts}
\usepackage{algorithmic}
\usepackage{graphicx}
\usepackage{textcomp}
\usepackage{xcolor}
\usepackage{multirow}
\usepackage{makecell}
\usepackage{todonotes}
\usepackage{subfig}

\interspeechcameraready

\title{DuRep: Dual-Mode Speech Representation Learning via ASR-Aware Distillation}

\author{Prabash Reddy}{Male}
\author{Swayambhu Nath}{Ray}
\author{Harish}{Arsikere}
\author{Akshat}{Jaiswal}
\author{Prakhar}{Swarup}
\author{Prantik}{Sen}
\author{Debmalya}{Chakrabarty}
\author{K V Vijay}{Girish}
\author{Nikhil}{Bhave}
\author{Frederick}{Weber}
\author{Sambuddha}{Bhattacharya}
\author{Sri}{Garimella}

\affiliation[nocounter]{Amazon AGI}{Bangalore}{India}
\email{prabash, swayar, arsikere, akshatkj @amazon.com}
\keywords{Speech encoder, dual-mode, best-in-class, HuBERT, Best-RQ, knowledge distillation.}

\usepackage{comment}

\begin{document}

\maketitle

\begin{abstract}
    
Recent advancements in speech encoders have drawn attention due to their integration with Large Language Models for various speech tasks. While most research has focused on either causal or full-context speech encoders, there's limited exploration to effectively handle both streaming and non-streaming applications, while achieving state-of-the-art performance. We introduce DuRep, a Dual-mode Speech Representation learning setup, which enables a single speech encoder to function efficiently in both offline and online modes without additional parameters or mode-specific adjustments, across downstream tasks. DuRep-200M, our 200M parameter dual-mode encoder, achieves 12\% and 11.6\% improvements in streaming and non-streaming modes, over baseline encoders on Multilingual ASR. Scaling this approach to 2B parameters, DuRep-2B sets new performance benchmarks across ASR and non-ASR tasks. Our analysis reveals interesting trade-offs between acoustic and semantic information across encoder layers.
\end{abstract}

\section{Introduction}
\label{sec:introduction}
The design of speech encoders, which extract meaningful representations from raw speech data, often involves a trade-off between performance and latency. Full-context encoders \cite{google_usm, whisper}, leveraging entire speech utterances, offer superior accuracy but higher latency, making them ideal for offline use. Conversely, streaming encoders process speech frame-by-frame, reducing latency for real-time applications like smart speakers and on-device assistants, though with some potential performance compromise. Recent efforts have focused on unifying full-context and streaming models for automatic speech recognition (ASR) tasks \cite{dual_mode_asr_google, dual_mode_conformer_nuance, multi_mode_stochastic_cmu}, combining the benefits of both encoder types while reducing training time and deployment costs, making them a compelling solution for speech-based applications.

Dual-mode frameworks proposed for ASR systems, such as \cite{dual_mode_asr_google}, unify streaming and non-streaming ASR models by jointly training both modes, assisted by an additional in-place distillation from full-context to streaming mode. However, this approach involves separate modules for convolution and normalization, leading to additional parameters. \cite{dual_mode_conformer_nuance} showed that model with shared convolution and normalization layers performs as good as models with mode-specific components.

Instead of limiting the model's context to full-context and pure streaming during training, \cite{multi_mode_stochastic_cmu} proposed sampling the future context from a distribution, adapting the model to a wide range of latency scenarios. To address the limitations of fixed-window masking and unintended look-ahead accumulation, \cite{dual_mode_conformer_nuance} divided input speech into non-overlapping chunks for attention. \cite{dual_mode_variable_apple} utilised variable attention mask to balance efficient modeling and reduced future frame dependency. Despite the focus on transducer \cite{rnnt} or CTC-based objectives, previous works have not explored distillation strategies for training dual-mode encoders, leaving a gap that our research aims to address.

In this paper, we present a zero-additional-parameter training framework for a dual-mode speech encoder adaptable to various latency scenarios. Our contributions include: 
\begin{itemize}
    \item A novel variable attention masking based distillation method to create a dual-mode encoder from a full-context teacher encoder.
    \item Analysis of the impact of varying teacher encoder and context sampling strategy on dual-mode encoder's performance.
    \item Comparative layer-wise analysis of semantic and acoustic feature representation using the SUPERB framework \cite{superb}.
    \item Demonstrate the scalability of our method by successfully scaling it to a 2B encoder, resulting in significant performance in both ASR and non-ASR tasks.
\end{itemize}

\section{Methodology}
\label{sec:methodology}

\subsection{Training stages}
\label{ssec:training_stages}

\begin{figure}[t]
  \centering
  \includegraphics[width=0.9\linewidth]{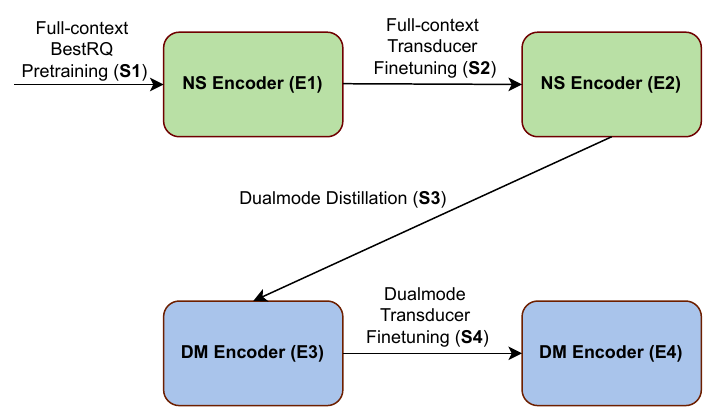}
  \caption{Dual-Mode encoder training framework}
    \label{training-overall}
\end{figure}

In this work, we propose a novel four-stage approach to develop a dual-mode encoder as demonstrated in Figure \ref{training-overall}:

\begin{itemize}
    \item \textbf{Full-context Pretraining (S1):} We begin by training a non-streaming (NS) encoder using BestRQ framework \cite{bestrq_google}, which is a widely accepted pretraining method in the industry for speech encoders on unlabelled data \cite{google_usm, assemblyai_universal1}. Pretraining helps prevent convergence issues for large model training.
    \item \textbf{Full-context Finetuning (S2):} The pretrained encoder from Stage 1 undergoes fine-tuning with an ASR-aware transducer objective, enhancing its performance for semantic tasks.
    \item \textbf{Dual-mode Distillation (S3):} A cross-entropy-based distillation process as outlined in Section \ref{ssec:knowledge_distillation}, is employed to transfer knowledge from the non-streaming ASR-aware encoder to a dual-mode (DM) encoder, aka ASR-aware distillation. This stage utilizes variable attention masking, as detailed in Section \ref{ssec:var_attn_masking}, to achieve seamless performance across both streaming and non-streaming modes.
    \item \textbf{Dual-mode Finetuning (S4):} Similar to Stage 2, the dual-mode encoder from Stage 3 is fine-tuned using an ASR-aware transducer objective with variable attention masking. This final stage further optimizes the encoder for dual-mode operation, resulting in a robust and versatile encoder for downstream tasks.
\end{itemize}

\subsection{Knowledge Distillation}
\label{ssec:knowledge_distillation}

Following the pseudo-target prediction approach outlined in \cite{hubert_1}, during S3 (shown in Figure \ref{training-overall}), we perform offline k-means clustering on the embeddings extracted from an intermediate conformer block of the full-context supervised encoder (E2). This block is chosen based on the semantic richness of its output representations (refer to section \ref{superb_layer_analysis}). The purpose of this clustering is to quantize the latent features of the full-context encoder, generating cluster IDs that serve as pseudo-labels for the dual-mode encoder (student) training. The dual-mode encoder is then trained to predict these cluster IDs. Our ablation studies showed that not applying masking to the input speech frames during distillation leads to a more effective student encoder.

\subsection{Variable Attention Masking}
\label{ssec:var_attn_masking}

In this work, we employ a variable attention masking strategy as discussed in \cite{dual_mode_variable_apple} to optimize how the encoder processes contextual information. We introduce \textit{variability} in the extent of the past and future contexts during the training process. Specifically, we define two uniform discrete distributions, $L_{past}$ and $L_{future}$, which encompass all possible values for past and future context, respectively. For each input batch, the look-back (LB) and look-ahead (LA) values are independently sampled from these distributions. This approach exposes the encoder to a range of contextual scenarios, improving its ability to function effectively under different latency conditions.

\begin{figure}[t]
  \centering
  \includegraphics[width=1.0\linewidth]{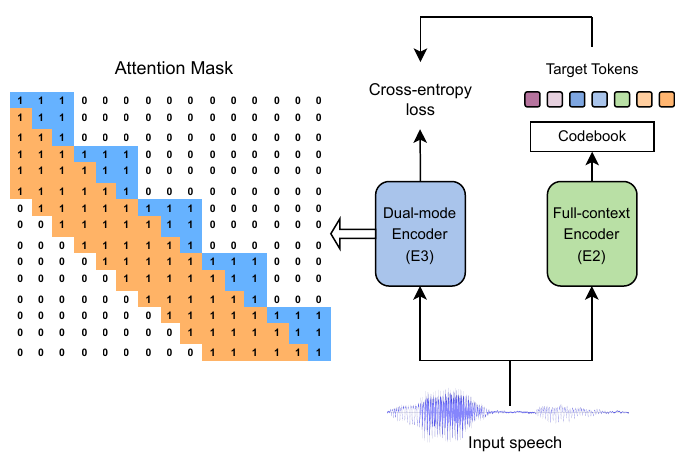}
  \caption{Cross entropy based distillation using variable attention masking - we use fixed-window for past and chunking for future context to prevent lookahead accumulation}
    \label{distillation-variable-masking}
\end{figure}

To prevent the unintended accumulation of look-ahead information across multiple encoder layers, we deliberately avoid applying fixed-window masking to future context. Specifically, we use chunked attention for future context, while a fixed-window approach is applied to handle past context. This strategy is visually represented in Figure \ref{distillation-variable-masking}, which illustrates a look-back of 5 frames and a look-ahead of 2 frames. The impact of distribution's sample space on encoder’s performance is examined in Section \ref{ssec:sample_dist_shift}.

\section{Experimental Details}
\label{sec:experimental_details}

We conducted all initial experiments and ablation studies on DuRep-200M, subsequently scaling our best-performing training strategy to DuRep-2B.

\subsection{Data}
\label{ssec:data}

We utilized a comprehensive speech dataset totalling 8.3M hours, which included both hand-transcribed speech data and a significant portion from untranscribed sources. Data consists of a combination of proprietary and publicly available datasets. Public datasets included Librispeech \cite{librispeech}, Multilingual Librispeech \cite{mls_dataset}, Commonvoice \cite{commonvoice_dataset}, PeopleSpeech \cite{peoplesspeech_dataset}, Fleurs \cite{fleurs_dataset}, VoxPopuli \cite{voxpopuli_dataset}, Librivox \cite{librivox_dataset} and others. For stages S1 and S3, which do not require transcriptions, we utilized the complete dataset.  For stages S2 and S4 (Transducer training), we used existing transcriptions where available and machine-generated transcriptions for the untranscribed partitions. These machine-generated transcriptions were produced using multiple proprietary models.


\subsection{Encoder Architecture}
\label{ssec:architecture}

Our proposed dual-mode encoder is based on the Conformer architecture \cite{conformer}, featuring a convolutional subsampling layer followed by a series of Conformer blocks. The input audio is processed into 128-dimensional log filter bank energy features, computed over a 25ms window with a 10ms shift. Each feature vector is stacked with 3 preceding frames and then downsampled to a 40ms frame rate, resulting in an overall input feature dimension of 512. The DuRep-200M model uses 1 convolutional front-end filter and 18 Conformer blocks, with each block having a model dimension (d\_model) of 512 and a feed-forward dimension (d\_ff) of 2048. Similarly, the DuRep-2B model also features 1 convolutional front-end filter and 20 Conformer blocks, with each block having a d\_model of 2048 and a d\_ff of 8192. All convolution operations are causal to prevent unintended look-ahead, and no positional embeddings are used, simplifying the encoder's overall design. In contrast to the original conformer paper, our conformer block places the convolution module before the multi-head self-attention module. This arrangement implicitly learns positional information without the need for additional positional embeddings.

After conducting a SUPERB layer analysis (\ref{superb_layer_analysis}) on E2, we selected the output from the 14th block for DuRep-200M and the 15th block for DuRep-2B for offline quantization during S3 (\ref{ssec:knowledge_distillation}). K-means clustering was performed on the extracted embeddings, with 2,500 clusters for DuRep-200M and 4,500 clusters for the DuRep-2B encoder. During transducer-finetuning (S2 and S4 of Figure \ref{training-overall}), the prediction network consisted of a 2-layer LSTM model \cite{lstm} with 1,280 hidden units, and the joint network contained 1,280 units. SpecAug \cite{specaug} is applied during all supervised training phases to enhance generalization. During variable-attention masking of S3 and S4, we sample past and future context (in seconds) uniformly from $L_{past}$=[$\infty$, 5.4, 4.6, 3.6] and $L_{future}$=[0, 1, 1.8, $\infty$]. These values were chosen based on our internal encoder usage scenarios.

\subsection{Baselines}
\label{ssec:baselines}
We adopt the well-established BestRQ framework for pre-training our 200M full-context and streaming baseline encoders. The pre-trained encoders are then individually fine-tuned using ASR transducer objective in their corresponding modes. Both models are trained until the convergence of Word Error Rate (WER) and the resulting fine-tuned encoders serve as single-mode baselines. The 200M baseline encoders are architecturally similar to the DuRep-200M encoder, with the exception that the non-streaming baseline encoder utilizes positional encoding and non-causal convolutions. During BestRQ pre-training, we mask 320ms of audio with a starting probability of 0.02 for the streaming baseline, and mask 400ms of audio with the same starting probability for non-streaming baseline.

\subsection{Evaluation Setup}
\label{ssec:evaluation_setup}

SUPERB \cite{superb, ml_superb} is a benchmarking framework designed to evaluate speech encoders and is used for relative comparison of encoders. During SUPERB evaluation, the candidate encoder is frozen, and a lightweight decoder (CTC for ASR and MLP for classification tasks) is trained. We evaluate our proposed encoders and all open-source encoders using an internal implementation of this framework that closely resembles the open-source SUPERB implementation. Major differences in the setup include: 1) we use a 4000 vocabulary wordpiece tokenizer for monolingual ASR and 15000 vocabulary wordpiece tokenizer for multilingual ASR tasks respectively instead of character based tokenizer, 2) instead of beam search we use greedy search for simplicity while evaluating CTC models and 3) no external LM is used in any evaluation. Due to these variations, absolute WERs reported after our SUPERB benchmarking are higher than those reported in the original works. But since the setup is kept constant across all runs, we can still use the WER numbers for relative comparisons across encoder models. We evaluate the encoders on Monolingual ASR (MonoASR), Multilingual ASR (MulASR), Language Identification (LID) and Emotion Recognition (ER) tasks. For simplicity, we report results on only ASR tasks for our 200M ablations.

MonoASR setup uses LibriSpeech corpus while MultiASR setup uses Multilingual LibriSpeech, CommonVoice, PeopleSpeech and VoxPopuli datasets during training and uses the test partitions of en, de, es, fr, it and pt of CommonVoice corpus for evaluation. For LID and ER, we use FLEURS and MSP Podcast \cite{msp_podcast} datasets respectively. Encoders are evaluated in Full-Context (LB, LA = ($\infty$, $\infty$)), Streaming (LB, LA = (5.4s, 0s)) and partial Look-Ahead (LB, LA = (5.4s, *)) modes.

\begin{table}[t]
\caption{Comparing DuRep-200M with single-mode baseline encoders using SUPERB on Monlingual and Multilingual ASR}
\centering
\setlength{\tabcolsep}{3.3pt}
\begin{tabular}{ccccccccccc}
\hline
\multirow{3}{*}{\textbf{Encoders}} & \multicolumn{2}{c}{\makecell{\textbf{Inference}\\\textbf{Setting}}} & \multicolumn{2}{c}{\textbf{MonoASR}} & \multicolumn{1}{c}{\textbf{MulASR}} \\
 & LB & LA & test-clean & test-other & test \\
 \hline
\multirow{4}{*}{Dualmode (E4)} & $\infty$ & $\infty$ & \textbf{4.61} & \textbf{10.16} & \textbf{21.25} \\ 
 & 5.4s & 1s & 5.12 & 11.30 & 23.35 \\
 & 5.4s & 0.6s & 5.25 & 11.91 & 24.43 \\
 & 5.4s & 0s & \textbf{5.81} & \textbf{13.63} & \textbf{27.31} \\
\hline
\makecell{Full-Context \\\ Baseline} & $\infty$ & $\infty$ & 5.07 & 11.36 & 24.05 \\
 \hline
\makecell{Streaming \\\ Baseline } & 5.4s & 0s & 6.75 & 15.70 & 31.06 \\
\hline
\end{tabular}
\label{results_200M}
\end{table}


\begin{table}[ht!]
\caption{Comparing ASR performance of DuRep-200M encoders obtained using different pre-training strategies}
\centering
\setlength{\tabcolsep}{3.3pt}
\begin{tabular}{cccccccc}
\hline
\multirow{2}{*}{\makecell{\textbf{Pre-Training} \\ \textbf{Strategy}}} & \multicolumn{2}{c}{\makecell{\textbf{Inference}\\\textbf{Setting}}} & \multicolumn{2}{c}{\textbf{MonoASR}} & \multicolumn{1}{c}{\textbf{MulASR}} \\
 & LB & LA & test-clean & test-other & test \\
\hline
\multirow{2}{*}{\makecell{BestRQ DM \\ Training}} & $\infty$ & $\infty$ & 6.024 & 13.13 & 25.29 \\
 & 5.4s & 0s & 6.851 & 15.76 & 31.69 \\
\hline
\multirow{2}{*}{\makecell{DM Distillation \\ from E1}} & $\infty$ & $\infty$ & 5.49 & 11.80 & 23.18 \\
 & 5.4s & 0s & 6.82 & 15.06 & 29.02 \\
\hline
\multirow{2}{*}{\makecell{DM Distillation \\ from E2}} & $\infty$ & $\infty$ & \textbf{4.61} & \textbf{10.16} & \textbf{21.25} \\ 
 & 5.4s & 0s & \textbf{5.81} & \textbf{13.63} & \textbf{27.31} \\
\hline
\end{tabular}
\label{dualmode_stages_ablation_table}
\end{table}

\section{DuRep-200M}
\label{sec:durep_200m}

\subsection{Results}
\label{ssec:durep_200m_results}

\begin{table}[ht!]
\caption{Effect of context's distribution during distillation (S3) on 200M dual-mode student encoder's (E3) performance}
\centering
\setlength{\tabcolsep}{3pt}
{\fontsize{8}{8.5}\selectfont
\begin{tabular}{cccccc}
\hline
\multirow{3}{*}{\makecell{\textbf{Sample Space}\\\textbf{Configurations} \\}} & \multicolumn{2}{c}{\makecell{\textbf{Inference}\\\textbf{Setting}\\\textbf{(seconds)}}} & \multicolumn{2}{c}{\textbf{MonoASR}} & \multicolumn{1}{c}{\textbf{MulASR}} \\
 \\
 & LB & LA & test-clean & test-other & test \\
\hline
\multirow{2}{*} {\textbf{T1}} & $\infty$ & $\infty$ & 6.02 & 13.42 & 26.75 \\
 & 5.4 & 1 & 6.39 & 14.54 & 29.38 \\
 \multirow{2}{*}{} & 5.4 & 0.6 & 6.64 & 15.12 & 30.28 \\
 & 5.4 & 0 & 7.65 & 17.36 & 34.56 \\
\hline
\multirow{2}{*}{\textbf{T2}} & $\infty$ & $\infty$ & \textbf{5.88} & \textbf{13.20} & \textbf{26.57} \\
 & 5.4 & 1 & 6.77 & 15.61 & 30.91 \\
 \multirow{2}{*}{} & 5.4 & 0.6 & 6.99 & 16.07 & 31.47 \\
 & 5.4 & 0 & \textbf{7.33} & \textbf{16.63} & \textbf{33.6}\\
\hline
\multirow{2}{*}{\textbf{T3}} & $\infty$ & $\infty$ & 6.14 & 13.45 & 26.63 \\
 & 5.4 & 1 & 6.38 & 14.85 & 29.14 \\
 \multirow{2}{*}{} & 5.4 & 0.6 & 6.64 & 15.03 & 30.06 \\
 & 5.4 & 0 & 7.62 & 17.4 & 34.87 \\
\hline
\end{tabular}
}
\label{sample_space_ablation_table}
\end{table}

Table \ref{results_200M} presents a comparative analysis of the dual-mode encoder and single-mode baseline encoders across full-context and streaming scenarios, in ASR tasks. In the full-context setting, Dual-mode encoder outperformed the baseline encoder, achieving an average improvement of 9.78\% and 11.6\% across monolingual and multilingual testsets. When evaluated in streaming mode, Dual-mode encoder shows an average performance gain of 15.14\% and 12\% on monolingual and multilingual testsets.

\subsection{Importance of 4-stage framework}
\label{ssec:importance_of_four_stage_framework}

\begin{table*}[ht!]
\caption{Comparison of DuRep-2B with open source models on ASR and Non-ASR tasks evaluated using SUPERB}
\centering
\begin{tabular}{cccccc}
\hline
\multirow{2}{*}{\textbf{Encoder (Inference Mode)}} & \multicolumn{2}{c}{\textbf{MonoASR}} & \multicolumn{1}{c}{\textbf{MulASR}} & \multicolumn{1}{c}{\textbf{Emotion}} & \multicolumn{1}{c}{\textbf{Language Id}}\\ 
 & test-clean(WER) & test-other(WER) & test(WER) &  MSP(Acc) & Fleurs(Acc)\\ 
\hline
DuRep-2B (FC) & \textbf{3.32} & \textbf{6.46} & \textbf{10.89} & \textbf{53.39} & \textbf{90.83}\\ 
DuRep-2B (S) & 4.32 & 9.27 & 15.51 & 52.73 & 88.74\\ 
\hline
Whisper-large-v3-encoder (FC) & 4.02 & 8.19 & 14.84 & 49.34 & 90.33\\ 
Wav2Vec2 1B (FC) & 6.05 & 11.97 & 35.24 & 49.66 & 36.85\\ 
Wav2Vec2 2B (FC) & 5.49 & 10.73 & 33.63 & 50.11 & 41.18\\ 
mHUBERT-147 95M (FC) & 5.22 & 10.77 & 27.85 & 50.31 & 87.9\\ 
WavLM 316M (FC) & 9.12 & 19.09 & 45.74 & 52.95 & 69.52\\ 

\hline
\end{tabular}
\label{encoder_comparison_2B_table}
\end{table*}

We conducted a series of experiments to understand the contribution of the first three stages in our proposed four-stage framework. In the first experiment, we pretrained a dual-mode encoder using the BestRQ framework. In the second experiment, we pre-trained a dual-mode encoder through dual-mode distillation from the unsupervised full-context encoder, E1. The third dual-mode encoder (E3), was prepared by distilling from the supervised full-context encoder (E2). All 3 encoders underwent dual-mode transducer finetuning (S4).

Results (Table \ref{dualmode_stages_ablation_table}), indicate that E4 outperforms the other encoders in both MonoASR and MultiASR tasks in both inference settings. This finding underscores the importance of distillation from a supervised full-context encoder in producing a high-quality pre-trained dual-mode encoder, which, when further fine-tuned, led to a superior dual-mode encoder.




\subsection{Role of context's sampling space during distillation}
\label{ssec:sample_dist_shift}

Table \ref{sample_space_ablation_table} summarizes SUPERB-ASR results for three different past and future context sampling configurations used during dual-mode distillation \textbf{(S3)}; \textbf{T1} = ($L_{past}$=[$\infty$, 5.4, 4.6, 3.6], $L_{future}$=[0, 1, 1.8, $\infty$]), \textbf{T2} = ($L_{past}$=[$\infty$, 5.4], $L_{future}$=[0, $\infty$]), \textbf{T3} = ($L_{past}$=[$\infty$, 5.8, 5.6, ..., 3.6], $L_{future}$=[0, 0.2, ..., 1.8, $\infty$]). The context values were chosen based on our internal encoder usage scenarios. T1 represents the ablation's base configuration. Restricting the sample space of encoder's context to either causal or full-context modes during training (T2) improved inference performance in full-context and causal modes by an average of 1.55\% and 3.72\%. However, it significantly degraded inference performance in partial look-ahead scenarios (by 6.17\% and 5.16\%). Expanding the sample space of encoder's context (T3) did not yield noticeable improvements for any inference scenarios. We attribute this to the inherent diversity of audio duration in the training data, which implicitly facilitates the model to attend to different contexts. Similar findings were observed during the dual-mode finetuning phase \textbf{(S4)}. As we aim to train an encoder that has reliable performance across deployment scenarios, we selected the base configuration (T1) for variable attention masking in S3 and S4.

\section{DuRep-2B}
\label{sec:durep_2b}


\subsection{Results}
\label{ssec:durep_2b_results}

Table \ref{encoder_comparison_2B_table} compares the performance of the DuRep-2B encoder in two inference modes with open-source encoders on ASR and non-ASR tasks, evaluated using SUPERB framework (\ref{ssec:evaluation_setup}). In full-context mode, the DuRep-2B encoder outperforms all encoders \cite{wav2vec2, mhubert, wavlm} across all tasks. It exceeds the performance of the whisper-large-v3 encoder by 19.26\% and 26.61\% on MonoASR and MulASR. Notably, the DuRep-2B encoder operating in purely causal mode, outperforms most open-source encoders operating in full-context mode across tasks.

\subsection{Conformer Layer Analysis Using SUPERB}
\label{superb_layer_analysis}

We conducted layer-wise analysis on encoder E2 and E4 using the SUPERB benchmarking setup to investigate the variations in acoustic and semantic information across the encoder layers. Unlike the standard SUPERB setup, which utilizes embeddings from all layers of the encoder, this analysis focused on embeddings extracted from individual encoder layers. We analyzed the encoder's semantic capabilities using Monolingual ASR and Speaking Rate Prediction, while Speech Intensity Contour Prediction and Pitch Contour Prediction were used to assess its acoustic capabilities using Dynamic Time Warping (DTW) with correlation distance measure. 

\begin{figure}[ht!]
  \centering
  \includegraphics[width=1.0\linewidth]{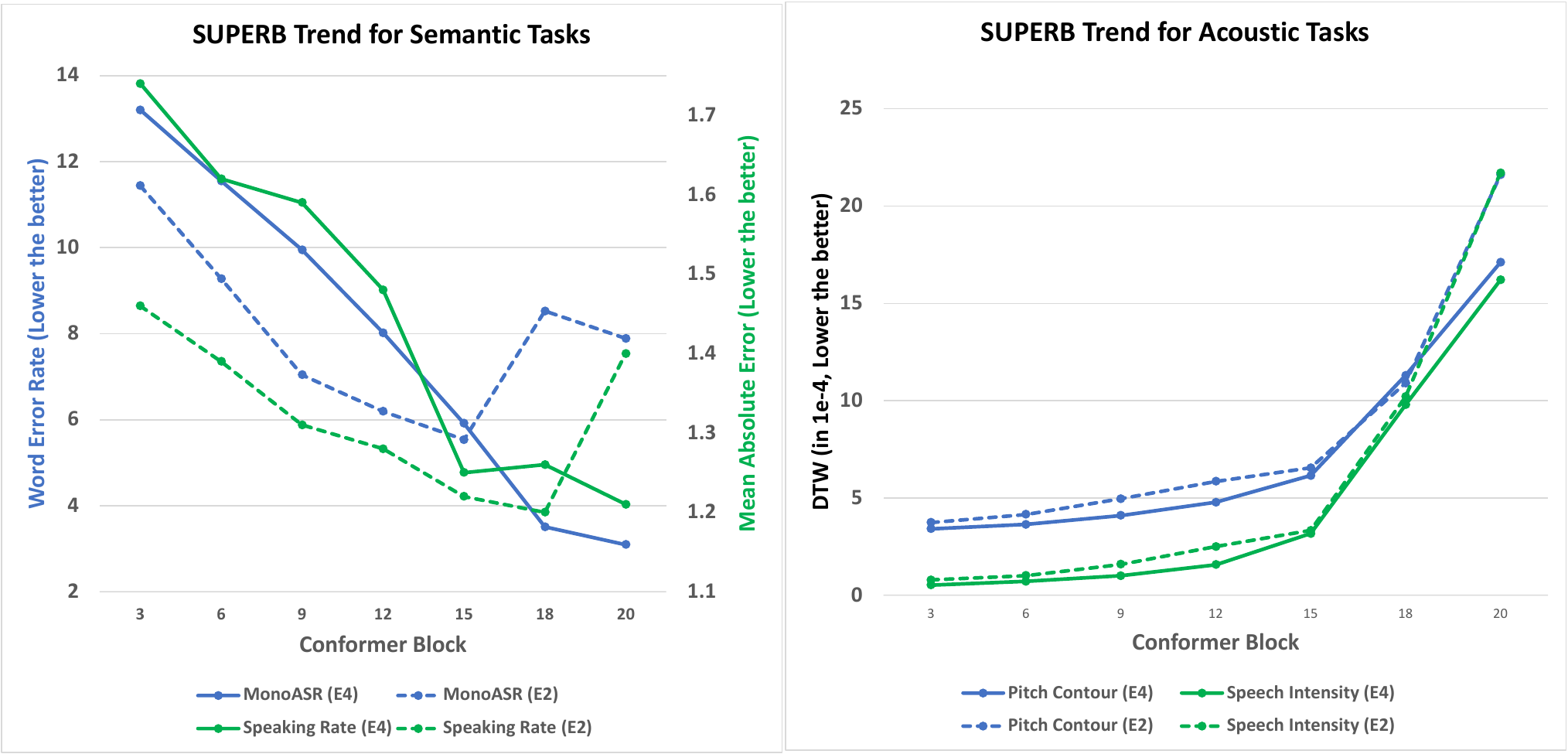}
  \caption{SUPERB style evaluation performed on semantic and acoustic tasks for 2B scaled versions of encoders E2 and E4}
    \label{superb}
\end{figure}

Figure \ref{superb} presents the development evaluation results for semantic and acoustic tasks across the layers of DuRep-2B. Our findings show that embeddings from the deeper Conformer blocks yield the best performance for semantically-oriented tasks like ASR and Speaking Rate Prediction. In contrast, for tasks that capture acoustic properties like Intensity and Pitch Contour Prediction, embeddings from the initial blocks of the encoder achieve optimal performance. \cite{lin2023utility} 
conducted a similar study on prosody reconstruction and related tasks using various self-supervised learning  models, showing that dominant prosodic information is mainly present in the initial layers.

Interestingly, our results reveal a unique trend for semantic tasks. The final Conformer block's embeddings provide the best performance for encoder E4, whereas embeddings from slightly earlier block yields the highest performance for encoder E2. We attribute the shift in peak performance towards deeper layers to the ASR-aware distillation stage (S3), which likely enhances model capacity utilization for semantic tasks.



\vspace{-.5cm}

\section{Conclusions}
\label{sec:conclusions}

We introduce DuRep, a dual-mode speech representation learning approach using variable attention based distillation, to train a general-purpose speech encoder that is optimized for multiple look-ahead inference scenarios. In ASR-SUPERB evaluations, on average, the DuRep-200M encoder outperforms baselines by 13.06\% in streaming and 10.42\% in non-streaming mode. The scaled-up DuRep-2B encoder shows average 21.71\% improvement over Whisper-v3 encoder in non-streaming mode. Despite not being fine-tuned for non-ASR tasks, DuRep-2B encoder outperforms open-source encoders across ASR and non-ASR tasks. Our analysis highlights the trade-off between semantic and acoustic information across encoder layers. 



\section{Acknowledgements}

We would additionally like to acknowledge Jahn Heymann, Ankish Bansal, Harish Mallidi, Phani Nidadavolu, Milind Rao, Bharat Padi, Venkata Kishore Nandury, Tuan Dinh, and Andreas Schwarz for their valuable contribution to data preparation efforts.

\bibliographystyle{IEEEtran}
\bibliography{Interspeech}

\end{document}